\documentclass[12pt]{article}
\usepackage{epsfig,latexsym}

\input epsf
\begin{document}
\font\cmss=cmss10 \font\cmsss=cmss10 at 7pt 
\hfill Bicocca-FT-00-22 \vskip .1in \hfill hep-th/0011041

\hfill

\vspace{20pt}

\begin{center}
{\Large \textbf{A NOTE ON REGULAR TYPE 0 SOLUTIONS}}
{\Large \textbf{AND CONFINING GAUGE THEORIES}}
\end{center}

\vspace{6pt}

\begin{center}
\textsl{F. Bigazzi$^{a}$, L. Girardello$^{a}$,  and A.
Zaffaroni$^{a,b}$} \vspace{20pt}

$^{a}$\textit{Universit\`{a} di Milano-Bicocca, Dipartimento di Fisica}

\textit{$^b$ INFN - Sezione di Milano, Italy}
\end{center}

\vspace{12pt}

\begin{center}
\textbf{Abstract }
\end{center}

\vspace{4pt} {\small \noindent We discuss some features of 
the regular supergravity solution for fractional 
branes on a deformed conifold, recently found by Klebanov-Strassler,
mostly adapting it to a 
type 0 non-supersymmetric context.
The non-supersymmetric gauge theory is $SU(M)\times SU(M)$ with
two bi-fundamental Weyl fermions.
The tachyon is now stabilized by the RR antisymmetric tensor flux.
We briefly discuss the most general non-supersymmetric theory
on electric, magnetic and fractional type 0 D3-branes on a conifold.
This includes the pure $SU(N)$ theory.}
\vfill\eject 
\noindent
\section{Introduction}
Supergravity solutions dual to non-conformal gauge theories 
have been extensively discussed in the past years. 
They have been obtained as type 0 solutions \cite{kt,kt1,minahan,k}, 
as deformations of conformal backgrounds
\cite{noi,freed,gub,flows,warner}
 or using fractional branes 
\cite{kn,ktfrac,ks,mn,tatar}. These backgrounds are dual
to strongly coupled gauge theories. The IR dynamics of such theories
can be different
from the familiar one of their weakly coupled cousins, but they are
nevertheless expected to exhibit familiar phenomena as confinement,
chiral symmetry breaking etc. Unfortunately, the IR region of the 
supergravity background, which should describe these phenomena, is
generically plagued with naked singularities. 

There are known examples of stringy resolution mechanisms.
The {\it enhancon} mechanism 
\cite{jpp} might serve for curing a generic class of 
repulson singularities,
while the expansion of dielectric branes into five-branes \cite{ps}
is adapt for describing some deformations of conformal theories.
An interesting solution that does not involve extra brane-like
sources and it is
completely regular has been proposed in 
\cite{ks}. It represents the dual of an 
$N=1$ gauge theory, realized by wrapping D5-branes on a 2-cycle of
a conifold.  Regularity is achieved by deforming the conifold
without introducing physical D3-brane sources. 

It is the purpose of this note to study the Klebanov-Strassler (KS)
solution \cite{ks}
in a non-supersymmetric context.  The natural candidates are  
type 0 solutions. The simplest examples are orbifolds
 of the KS solution. 
Since type 0B is an orbifold of type IIB, every type IIB brane
solution leads to a type 0B solution for orbifolded branes in the
regular representation. The $AdS_5\times S^5$ dyonic solution
of \cite{kt1} is an example of this correspondence \cite{ns}.

The tachyon instability of type 0 needs to be cured.
Fortunately, for this class of regular solutions the type 0 tachyon is 
 stabilized by the RR antisymmetric tensor flux.
As in \cite{kt,kt1}, the stabilization mechanism is effective 
only for large or intermediate curvature. In this regime,
we expect to recover the string dual of
weakly coupled gauge theories. We can speculate that at large t'Hooft
coupling a large N phase transition has occurred \cite{k},
making the theory unstable. The
supergravity solution nevertheless is a good starting point 
for the formulation of the string model problem and
for a qualitative analysis of the dynamics.
 
We will also discuss more general solutions 
with running tachyon and dilaton, corresponding to the non-regular
representation. These are the analogous of the purely electric 
type 0 solutions discussed in \cite{kt}.  
Most of the gauge theories
discussed in this paper have no massless scalar fields, thus avoiding
some complications in \cite{kt}. We start an analysis 
of the type 0 dual of the pure $SU(M)$ gauge theory. We mostly
focus on the IR behavior. Under the assumption that the tachyon 
relaxes to zero and its instability is cured by the RR flux,
we find that the KS solution can be easily adapted
to the IR description of the pure glue theory. We work at large N.
We may expect corrections to the gauge theory behavior already
at the first order in $1/N$ \cite{armoni}.

\section{The KS solution}
Let us start with a brief review of the type IIB solution.
There is a class of exact type IIB solutions \cite{pgrana,cvetic},
that do not necessarily request supersymmetry.
Consider the general class of black D3-brane solutions
\begin{eqnarray}
ds^2&=&Z^{-1/2}dx_{\mu}dx^{\mu}+Z^{1/2}ds^2_K\nonumber\\
F_5&=&da_4+*da_4,\qquad a_4={1\over 4Z}dx^0\wedge dx^1\wedge dx^2 
\wedge dx^3 
\label{black}
\end{eqnarray}
where $K$ is a Calabi-Yau manifold with a closed self-dual 3-form   
$\omega_3$, $*_K\omega_3=i\omega_3$. If the 3-form 
$G=H_{(3)}+ig_sF_{(3)}$ is proportional to
$\omega_3$, its equation of motion is
automatically satisfied and we obtain a class of exact solutions
by imposing
\begin{equation}
- \Box_K Z=\rho_{D3}(x) + G_{mnp}G^{(K) mnp}
\label{eqZ}
\end{equation}
where $\rho(x)$ is a general density of D3-branes \cite{pgrana}. 
Notice that 
the Laplacian and the metric in $G^2$ are those of $K$.

Regular and fractional branes on a conifold nicely fit into eq. 
(\ref{black}), as noted in \cite{pgrana,gub2}. 
Let us briefly discuss some details of the solution
\cite{kg,kn,kt,ks}. N D3-branes on a conifold give rise to
a conformal theory with $SU(N)\times SU(N)$ gauge group, 
whose supergravity dual is $AdS_5\times T^{1,1}$
\cite{kw}. Since a D5-brane wrapped on a 2-cycle of $T^{1,1}$
(fractional brane) carries D3-charge, we have a system with two
types of D3-branes. The gauge theory corresponding to N physical
 and M fractional D3-branes is $SU(N)\times SU(N+M)$ \cite{kw,kg,kn}.
In the $AdS_5\times T^{1,1}$ background, we
turn on M units of RR-flux on $T^{1,1}$. 
The solution was originally found by enforcing 
supersymmetric-inspired first order equations of motion \cite{kt}.
Remarkably, this solution
satisfies all the requests for the class of exact solutions (\ref{black})
\cite{pgrana}.
Supersymmetry or imposition of the black D3-brane ansatz (\ref{black})
are extremely useful for solving
the equations of motion of type IIB supergravity, but they do not
guarantee by themselves regularity of the resulting solution.
The  $SU(N)\times SU(N+M)$ theory flows to the IR region by
forgetting shells of regular branes via an enhancon mechanism
\cite{jpp,ktfrac,ks}.
In the far IR, the dynamics is that of $SU(M)$ SYM without 
regular branes and the
background is made regular by the hypothesis that the conifold
is replaced by a deformed conifold \cite{ks}. The explicit
solution can be found in \cite{ks} and the asymptotic IR
behavior is reviewed in Section 4. We just quote the behavior
of the conifold metric and the Z function in the IR (small $\tau$).
\begin{eqnarray}
ds^2_K&=&{d\tau^2\over 2}+{\tau^2\over 4}d\Omega_2^2+d\Omega_3^2
\nonumber\\
Z&\sim&(g_sM)^2+O(\tau^2)
\label{defsol}
\end{eqnarray}
In the deformed conifold at small $\tau$ the 2-cycle is collapsed
but it remains a finite three-sphere
of radius squared $O(g_sM)$, supporting the RR flux.
The complex field $G_{(3)}=H_{(3)}+ig_sF_{(3)}$  is self-dual $*_K G_{3)}=
iG_{(3)}$, and the R-R and NS-NS three-forms satisfy 
\begin{equation}
g_s^2F^2_{(3)}=H^2_{(3)}.
\label{relation}
\end{equation}
 $g_sM$ plays
the role of the gauge theory t'Hooft parameter and it has to be large
for the supergravity approximation to be trusted.

Notice also that another regular supersymmetric solution for $N=1$ SYM
appears in \cite{mn}. The crucial ingredients, a vanishing 2-cycle
and a finite-size 3-cycle supporting the RR flux, are the same.  

The class of exact solutions (\ref{black}) does not require
supersymmetry. 
We want to study the solutions (\ref{black}) in the context 
of type 0B.
\section{Type 0 branes on a conifold}

Let us start with a general discussion of the theory of 3-branes
on a conifold with general charges.

It is useful to think about type 0B as an orbifold
of type IIB by the spacetime fermion number. The untwisted
sector of the orbifold contains all the bosonic fields of type IIB,
which we denote as 
$g_{\mu\nu}^{UT},B_{\mu\nu}^{UT},\phi^{UT},F_{(n)}^{UT}$, 
while the twisted sector contains
a NS-NS tachyon $T$ and a new set of R-R fields $F_{(n)}^{T}$.
The tachyon and R-R part of type 0B Lagrangian is 
better formulated in terms of the
combinations $F_{(n)}^{\pm}=(F_{(n)}^{UT}\pm F_{(n)}^T)$
\footnote{It is possible that a generic
function $g(T)=1+\alpha T^2+...$ multiplies the NS-NS antisymmetric
field kinetic term. It has been explicitely checked \cite{tseytlinproc}
that $\alpha=0$. Higher terms in $T$ do not affect our conclusion
and, therefore, we set $g(T)=0$ in the following.}   
\cite{kt}
\begin{equation}
L=\int d^{10}x \sqrt{g}\left ( -2e^{-2\phi}T^2+
\sum_{n=1,3,\pm} {1\over 2}
(1\pm T+{T^2\over 2})|F_{(n)}^{\pm}|^2+(1+T+{T^2\over 2})
|F_{(5)}|^2\right )
\label{lag}
\end{equation}
The Lagrangian is written in string frame and is expanded
to quadratic order in T. $F_{(5)}$ has now both self-dual 
or antiself-dual components and
appears in the Lagrangian as the electric variable. 
Bianchi identities now read
\begin{equation}
dF_{(5)}=H_{(3)}\wedge F_{(3)}^+,\,\,\,\,\,\,\,
d*F_{(5)}=H_{(3)}\wedge F_{(3)}^-
\label{bianchi}
\end{equation}

Since all the R-R fields are doubled, also D-brane are. On a conifold,
we now have a total of 4 different 3-brane charges, electric and
magnetic D3 charges \cite{kt} and wrapped D5$^+$ and D5$^-$ 
charges\footnote{An electric or magnetic D3-brane 
in type 0 can be considered
as a fractional brane of the $(-1)^F$ orbifold. To avoid confusion,
we reserve the name {\it fractional} for the wrapped D5-branes.}.

\begin{figure}
\centerline{\epsfig{figure=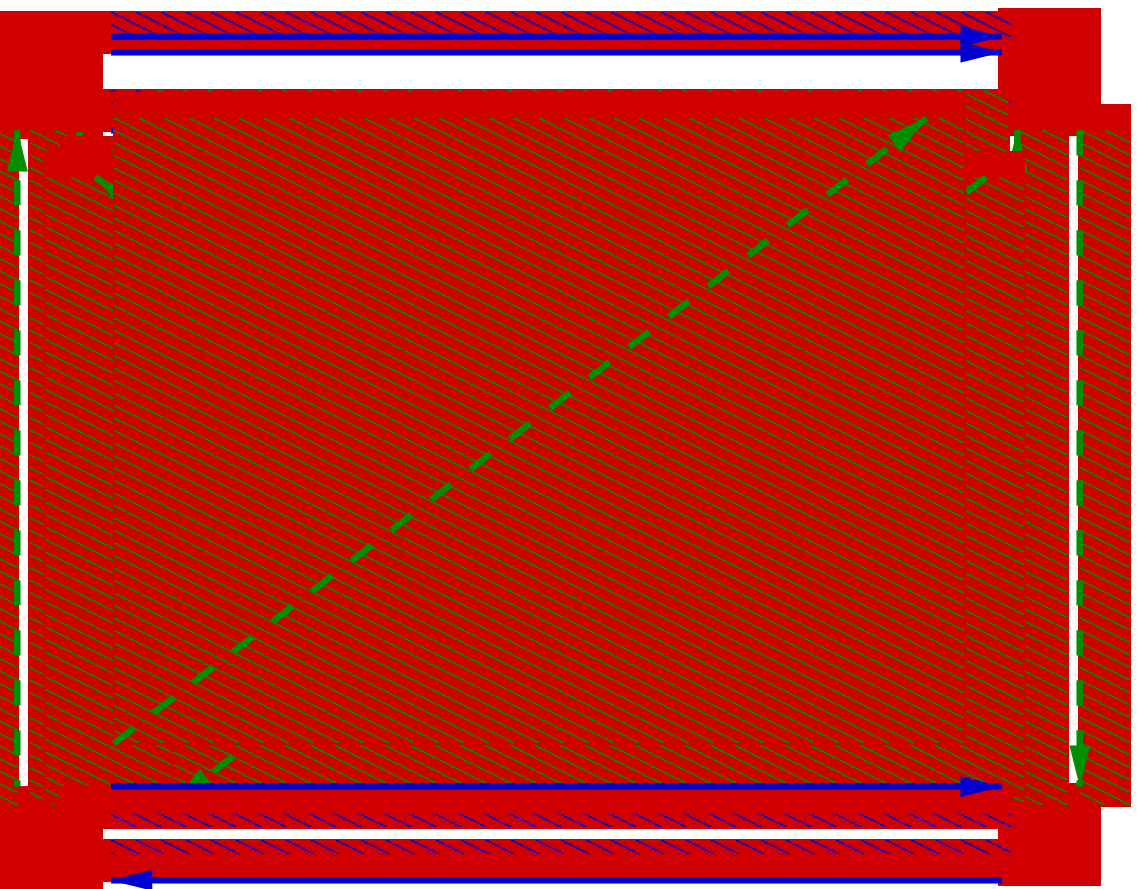,height=5 cm, width=6cm}}
\caption{The generic gauge theory for type 0 branes on a conifold.
The nodes are $SU(N_i)$ gauge groups. The links represent bi-fundamental
matter fields. Solid lines are complex bosons and dashed lines 
Weyl fermions. Arrows distinguish between $(N_i,\bar N_j)$ 
and $(N_j,\bar N_i)$ representations.}
\end{figure}
The generic theory containing all the four types of 3-brane charges
can be determined as in \cite{kt,scarafaggio,luest}. It
is $\prod_{i=1}^4SU(N_i)$ with the matter content indicated
in the quiver diagram of Figure 1. There is in addition 
a complicated non-renormalizable potential for the matter field
obtained from the projection of the superpotential in \cite{kw}.

Incidentally, we notice that, as in type II \cite{uranga}, 
the brane configuration has a type 0A 
T-dual description consisting of D4$^\pm$ and NS branes.
The NS world-volume is, say, along $(0,1,2,3,4,5)$ and $(0,1,2,3,8,9)$
and the NS branes
 are at the position $0,x_6^{(0)}$ in the compact direction
$x_6$. There are $N_1,N_2$ D4$^\pm$ branes stretching in $x_6$
between the NS branes along $(0,x_6^{(0)} )$ and $N_3,N_4$ 
D4$^\pm$ branes along $(x_6^{(0)},2\pi R_6)$. Standard rules for
brane configurations, adapted to type 0 \footnote{Some
example of similar
type 0 brane configurations can be found in \cite{kol}.},
 reproduce the gauge theory
and matter content in Figure 1. 

We are also interested in studying the deformed conifold.
Let us introduce an ansatz for the type 0 solution along the
lines of ref. \cite{ks}. The metric ansatz describes both a conifold
and a deformed conifold. The antisymmetric tensor ansatz is the most
general one compatible with the $Z_2$ conifold symmetry and
the expected RR flux\footnote{The RR flux is supported by the conifold
3-cycle in the UV and by the $S^3$ that replaces the singularity
in the deformed conifold  in the IR.}.
The 10 dimensional solution is 
\begin{eqnarray}
ds^2&=&e^{-5q}(dr^2+e^{2A(r)}dx_{\mu}dx^{\mu})
+ds^2_{5\prime}\nonumber\\
ds^2_{5\prime}&=&e^{3q-8f}g_5^2+e^{3q+2f+y}(g_1^2+g_2^2)+
e^{3q+2f-y}(g_3^2+g_4^2)\nonumber\\
B_{(2)}&=&g(r)g_1\wedge g_2+k(r) g_3\wedge g_4\nonumber\\
F_{(3)}^{\pm}&=&M_{\pm}g_5\wedge g_3\wedge g_4+d[F(r)(g_1\wedge g_3+
g_2\wedge g_4)]\nonumber\\
F_{(5)}&=& K_+(r)g_1\wedge g_2\wedge g_3\wedge g_4\wedge g_5
+ *(K_-(r)g_1\wedge g_2\wedge g_3\wedge g_4\wedge g_5)
\label{10}
\end{eqnarray}
where $K_{\pm}=N_{\pm}+(k-g)F_{\pm}+2M_{\pm}g$.
Explicit expressions for the $g_i$ forms can be found in
\cite{ks}. Notice that, while the Bianchi identities for the 5-form
are automatically satisfied, no self-duality condition is
imposed on the 3-forms. The requirement of self-duality
in the class of solutions (\ref{black}) amounts to a 
precise relation between $Z$ and the unknown functions in the
3-form ansatz. In \cite{ks}, this relation is imposed by requesting
first order equations. From the point of view of \cite{pgrana},
it is not so strictly related to supersymmetry but it is a part
of the exact solution (\ref{black}). The field $y$ 
allows for deformations of the conifold.
$N_{\pm}$  
are the number of electric and
magnetic physical D3 branes and $M_{\pm}$  are the number
of fractional branes.
With these notations the gauge group is $SU(N_+)\times SU(N_-)\times
SU(N_+ +M_+)\times SU(N_- +M_-)$.

Using the results in 
\cite{kt,ktfrac,tseytlin}, it is easy to write a 
five-dimensional  effective action for the relevant fields,
\begin{eqnarray}
L&=&\int dr e^{4A}\left ( 3(\partial A)^2-{1\over 2}G_{ab}\partial\phi^a
\partial\phi^b -V(\phi )\right )\nonumber\\
G_{ab}\partial\phi^a\phi^b&=&15(\partial q)^2+10(\partial f)^2
+{(\partial T)^2\over 8}+ {(\partial y)^2\over 2}+
{(\partial\phi)^2\over 4}\nonumber\\
&+&e^{-\phi-6q-4f}
(e^{-2y}{\sqrt{3}(\partial g)^2\over 2}+e^{2y}{\sqrt{3}(\partial k)^2\over 2})+
\sqrt{3}\sum_{\pm}f_{\pm}(T)e^{\phi 
-6q-4f}{(\partial F_{\pm})^2\over 2}\nonumber\\
V(\phi)&=&e^{-8q}\left (e^{-12f}-6e^{-2f}\cosh{y}+{9\over 4}e^{8f}
\sinh{y}^2\right )-{T^2\over 8}e^{\phi/2-5q}\nonumber\\
&+&\sum_{\pm}\left [ {9\sqrt{3}\over 8}e^{4f-14q+\phi}f_{\pm}(T)
(e^{-2y}F_{\pm}^2+e^{2y}(2M_{\pm}-F_{\pm})^2)
+{27\over 4}e^{-20q}f_{\pm}(T)
K_{\pm}^2\right ]\nonumber\\
&+& {9\sqrt{3}\over 8}e^{4f-14q}
e^{-\phi}(g-k)^2
\label{pot}
\end{eqnarray}
where $f_{\pm}=1\pm T+T^2/2+...$. Einstein equations also impose the
constraint $3(\partial A)^2-{1\over 2}G_{ab}\partial\phi^a
\phi^b +V(\phi )=0$.
The undeformed conifold is obtained for $y=0,k=g,F_{(\pm)}=0$.

We analyze the potential in various steps.
First, set $M_{\pm}=0, N_-=N_+=N$ and $T=0$ and identify
all $\pm$ fields: $\lambda_-=\lambda_+$.
 The potential is now indistinguishable
from the potential for N type IIB D3-branes at a conifold.
By rescaling fields, we can set $N=-2/3\sqrt{3}$. 
Consider first the type IIB potential.
For $q=f=y=g=k=F=0$ 
we have an $N=1$ critical point corresponding
to $AdS_5\times T^{1,1}$. From eq. (\ref{pot}), we can extract
the mass of the fields and the dimensions of the corresponding 
operators: $\Delta_q=8,\Delta_f=6,\Delta_y=3,\Delta_{k-g,F}=3,7$.
$k+g$ does not appear in the potential. It is the massless
field $\int_{S^2}B_{(2)}$ associated with a
marginal direction in the CFT \cite{kw,kw2,kn}. 
 The corresponding operator is 
${1\over g^2}(F^2_{(1)}-F^2_{(2)})$. 
Using the results in \cite{ferrara},
we can tentatively identify these fields with operators in the
multiplets:
\begin{eqnarray}
&&q,f\rightarrow {\mbox Tr} (W^2\bar W^2)\nonumber\\
&&(k-g,F)_{\Delta =7}\rightarrow  
 {\mbox Tr}(A\bar A+B\bar B)W^2\nonumber\\
&&(k-g,F)_{\Delta=3}\rightarrow {\mbox Tr} (W^2_{(1)}+W^2_{(2)}) 
\nonumber\\
&&y\rightarrow {\mbox Tr} (W^2_{(1)}-W^2_{(2)})
\label{op}
\end{eqnarray}

The terms proportional
to $M_{\pm}$ come from the introduction of 
wrapped D5-branes. If we still set $\lambda_-=\lambda_+$ and $T=0$
we find the potential for the $N=1$ configuration of regular and 
fractional type IIB branes \cite{ktfrac}.
Finally, the type 0 reduction is accomplished by turning on the
tachyon field $T$ and by doubling all the fields associated with
RR forms.  
\section{A conformal solution}
We can consider configurations of D3 and wrapped D5-branes in
type 0B as projections of similar configurations in type IIB.
The simplest configuration is obtained by using the regular
representation in the orbifold projection \cite{dm}. Every
type IIB brane is doubled. The $AdS_5\times S^5$ dyonic solution
of \cite{kt1} is an example of this procedure \cite{ns}.
The type 0B solution is obtained by identifying the untwisted
fields with their parents in type IIB and setting to zero all
the twisted fields.   

We expect a non-supersymmetric conformal solution as a projection
of the $N=1$
$AdS_5\times T^{1,1}$ background. This solution is 
also discussed in \cite{scarafaggio,luest}. It does not contain 
fractional branes. Every IIB D3-brane
is now doubled into an electric and magnetic type 0
D3-brane. The resulting gauge theory is
$SU(N)^4$ with the matter content indicated in figure 1
\cite{scarafaggio,luest} and a non-renormalizable potential. 
We expect the theory to flow in the IR to
an interacting fixed point as in \cite{kw}.

It is easy to verify that for $M_{\pm}=0$ and $N_-=N_+=N$, 
$q=f=T=\phi =F_{\pm}=g=k=0$ is an $N=0$ critical point of the potential
(\ref{pot}) corresponding to a 10 dimensional solution
$AdS_5\times T^{1,1}$. 

There are a total of four independent coupling constants
in the gauge theory. We can identify the real part of three of these
couplings with the dilaton, the value of $B_{(2)}$ on the 2-cycle and the
tachyon. The constant value of the dilaton and  $B_{(2)}$
do not appear explicitly in the 10 dimensional equations of motion.
There are therefore two exactly marginal direction in the CFT as in 
\cite{kw}, at least in the large N limit.
The existence of such marginal directions in a non-supersymmetric
theory is easily explained.
In the large N limit, the correlation functions for untwisted 
fields are  identical 
to those in the parent type IIB theory \cite{silverstein}. 
However, we do not expect that the coupling constant 
associated with the tachyon, which is a twisted field, 
is a marginal parameter. 
The dimension of the corresponding 
operator is computed below. It would be interesting to identify
the fourth coupling within the supergravity modes. 

The spectrum of operator dimensions can be easily extracted from
eq. (\ref{pot}).
The fields considered before eq. (\ref{op}) are still in the
spectrum with the same mass as in type IIB. 
The type 0 operators are obtained by replacing fermions/scalars with
bi-fundamental fermions/scalars and considering
particular combinations
of the traces in the four gauge groups.
We have two more fields from the type 0 twisted sector,
the tachyon $T$  and a combination $t$ of $(F_{(\pm)},k,g)$.
The corresponding operators have mass-squared
\begin{eqnarray}
T&\rightarrow& m^2= 16-2\sqrt{x}\nonumber\\
t&\rightarrow& m^2= 9,\,\,\, \Delta =2+\sqrt{13}
\label{twisted}
\end{eqnarray}
where $x$ is the t'Hooft coupling of the overall coupling constant
\footnote{Notice that we rescaled a factor of $N$, so that $x=e^{\phi}$.}.
As in \cite{k}, the tachyon is unstable at large $x$. It is tempting
to speculate as in \cite{k} that the theory is perfectly unitary
at small $x$ (as the extrapolation of formula (\ref{twisted})
indicates) and undergoes a phase transition at larger $x$, as typical
of many gauge theories in the large N limit.
  
\section{A non-conformal ``self-dual'' solution}
The next example is obtained by introducing D5$^\pm$ branes
but still enforcing an orbifold projection with the regular 
representation.  
The $N=1$ solution for a $SU(N+M)\times SU(N)$ gauge theory 
\cite{ktfrac,ks} has been briefly reviewed in the Introduction.
It descends to a non-supersymmetric 
type 0 solution by putting 
$T=0$ and $\lambda_-=\lambda_+$ for all the relevant fields 
and parameters.
The gauge theory is $SU(N)^2\times SU(N+M)^2$. We have N self-dual
D3-branes, M wrapped D5$^+$ and M wrapped D5$^-$.
The supersymmetric solution flows 
in the IR to the pure $N=1$ SYM with a cascade mechanism \cite{ks}.
The corresponding IR solution contains no D3-branes
and is defined on a deformed conifold \cite{ks}.
We expect that 
the cascade mechanism has a non-supersymmetric analogous. 
The gauge theory dynamics is obviously much more difficult to analyze
due to the lack of supersymmetry. However, we may expect that,
in the large N limit, many supersymmetric results based on duality
should continue to hold \cite{silverstein, schmaltz}.
Notice that 
there are no reasons for imposing first order equations in a non
supersymmetric context.
The cascade solution is just one particular solution of type 0
out of infinitely many. The fact that it is completely regular
indicates that it actually corresponds to a possible
quantum field theory RG flow.

The IR gauge theory is an $SU(M)\times SU(M)$ gauge theory
with 2 bi-fundamental Weyl fermions. The IR dynamics
of this theory is described by the type 0 restriction of the
deformed conifold solution. For completeness, we quote from \cite{ks}
the IR behavior of the solution (small $\tau$).
\begin{eqnarray}
ds^2_K&=&{d\tau^2\over 2}+{\tau^2\over 4}d\Omega_2^2+d\Omega_3^2
\nonumber\\
Z&\sim&(g_sM)^2+ O(\tau^2)\nonumber\\
g&\sim& \tau^3,\,\,\,\, k\sim\tau,\,\,\,\, F\sim \tau^2
\label{IR}
\end{eqnarray}
The complete solution can be found in \cite{ks}. It is also
derived in \cite{tseytlin} using the 5 dimensional formalism of eq. (\ref{pot}). The radial coordinate $\tau$ is related to the coordinate $r$ 
of the five-dimensional effective theory by $d\tau =-e^{4f-4q}dr$.
Notice that in the five-dimensional approach, the metric
fields $q,f,y,A$ are singular, behaving as $\sim \log (r)$.
The five-dimensional metric has therefore a naked singularity
$ds^2\sim dr^2+r^{4/5}dx_\mu dx^\mu$. This is just an artifact of
the dimensional reduction, the ten-dimensional solution being
completely regular for all $r$.
  
The solution exhibits confinement and spontaneous chiral symmetry
breaking \cite{ks}. Incidentally, we notice that the deformation of the
conifold is requested for studying vacua with non-zero
fermionic condensate.  
We discuss the type IIB solution,
all considerations being applicable to type 0 too. 
We use standard techniques \cite{noi,freed,rev}, with a proviso. 
Since the gauge theory is not a deformation of a CFT, the
supergravity solution is not asymptotic to a purely AdS vacuum.
We can however consider a regime, for example $M<<N$, where
modifications of the AdS solution are small. The space has still
a boundary and we can reasonably use the machinery of AdS/CFT
for identifying operators. Adapting the results in \cite{tseytlin},
we can write a superpotential \cite{freed} for the type IIB
supersymmetric theory
\begin{equation}
W=-3e^{4f-4q}\cosh y -2e^{-6f-4q}-3\sqrt{3}e^{-10q}\left (
N+F(k-g)+2Mg\right )
\label{superpo}
\end{equation} 
The tadpole term $2Mg$, which is the same for a conifold or a deformed
conifold, is due to the fractional branes. This
term is responsible for the change of  gauge group
$SU(N)\rightarrow SU(N+M)$. The running of fields due to the
remaining terms in the superpotential can be interpreted as 
a deformation or a change of vacuum in the gauge theory \cite{bala,kw2}.
Substitution of the conifold with the deformed
conifold introduces the fields $y,F,(k-g)$. According to
eq. (\ref{op}), these fields correspond to the gaugino condensates
of the two gauge groups. 
We can decide whether the running of these fields corresponds to
a deformation or a choice of a different vacuum by looking 
at the asymptotic behavior \cite{bala,kw2,rev}.
The expansion of W near the origin determines the asymptotic behavior
using $2G^{ab}\dot\lambda_b=\partial W/\partial\lambda_a$.
For canonically normalized fields $\lambda_C$, 
$W=-3+\alpha \lambda_C^2+...$ implies $\lambda_C\sim e^{-\alpha r}$.
$\alpha=\Delta$ corresponds to a vacuum of the theory,  $\alpha=4-\Delta$ 
to a deformation. In our case, for canonically normalized fields,
\begin{equation}
W\sim -3 +4q^2_C-6f^2_C-3y^2_C-6F_C(k-g)_C+...
\label{Was}
\end{equation}
We can easily see that the fields associated 
with the gaugino condensates, $y_C$, and a combination of $F_C,(k-g)_C$
behave as $e^{-\Delta r}=e^{-3r}$.
The corresponding operators have non-zero vacuum expectation value. 
Therefore, one motivation for introducing the deformation of the
conifold is to study vacua with non-zero
gaugino condensate, as it is the case for $N=1$ SYM. Similar results
apply to the fermionic condensates of type 0 solutions.

There is still a fundamental question. The tachyon vacuum
expectation value is zero. Is the background stable?
In type 0 models, the stabilization of the tachyon is due to the
D3-brane charge, which is absent here. Nevertheless, we
have an antisymmetric tensor background that may serve the purpose.
Let us qualitatively investigate the effective mass for the tachyon
field. In our background, since $F^{T}=0$, we
have $F_{(3)}^+ =F_{(3)}^-=H_{(3)}/g_s$ and the effective 
potential for the tachyon follows from eq. (\ref{lag})
\begin{equation}
L=-2e^{-2\phi}T^2+ (1+{T^2\over 2})|F_{(3)}^+|^2
\label{stabilization}
\end{equation}
From this equation we see that, as anticipated, $T$ has no tadpoles
and can be consistently set to zero. Moreover, its effective mass
is \footnote{Notice that the NS-NS two-form does not contribute to
the tachyon mass \cite{tseytlinproc}.}
\begin{equation}
m_T^2\sim -{2\over g_s^2}+ {M^2\over 2(g_sM)^3}
\label{mass}
\end{equation}
where we used the fact that the $S^3$ has radius-square $(g_sM)$.
We see that the stabilization condition is $-2+{1\over 2g_sM}\ge 0$.

If $g_sM$ is sufficiently small, the background is stable.
Notice that this is the limit where supergravity 
is no longer valid. The same
happens in the known type 0 solutions \cite{kt,kt1}.
The tachyon instability may correspond to some pathological
behavior for the strongly coupled gauge theory. The description
of the weakly coupled $SU(M)^2$ theory with bi-fundamental fermions
requires small $g_sM$ \footnote{The focus into the IR by decoupling
 the UV region and the infinite cascade needs
small $g_sM$ \cite{ks}.}.  
The supergravity approximation is then invalidated
and a string sigma-model analysis is requested.  
Extrapolating from eq. 
(\ref{mass}), we may conclude that the associated string 
background is stable. 
We can image a situation \cite{kt1}
where the stabilization mechanism, made more precise,
just requires an intermediate (of order 100) value for the
t'Hooft parameter and corrections can be still estimated to be small.
Alternatively, we should request the knowledge of the
string sigma-model. 
The absence of $F_{(5)}$ background may simplify the problem. 

\section{The pure glue theory}
We now turn to the system with
only one type of fractional brane, say D5$^+$. This is the most
interesting theory, since it is pure glue $SU(M)$. 

A solution for D5$^+$ branes 
corresponds to the non-regular representation in the orbifold reduction
of the type IIB solution. An analogous configuration in type 0,
consisting of purely electric D3 branes with gauge group $SU(N)$ and
6 massless bosons, was considered in \cite{kt}. 
By avoiding the existence of massless bosons, we also avoid all
the related complications and fine tuning problems \cite{kt,minahan}.
The IR behavior of
the equations of motion for electric D3-branes
was considered in \cite{minahan,kt,maggiore} and, as expected,
is generically singular. 
Such asymptotic behavior contains very few information about what
kind of solution is really selected by physics and how 
the singularity is resolved by 
dynamics.

Here we propose a possible regular background describing the IR dynamics
of the pure glue theory. Once again,
we look for regular solutions on a
deformed conifold. Since $F_{(3)}^-=0$, we see from eq. (\ref{lag})
that the tachyon has now a tadpole and has to run.
It may induce a running of the dilaton.

We assume that the tachyon relaxes to zero in the IR. 
Most likely, this is a fine-tuning.
We also make the assumption
that the dilaton goes to a constant value.  
All the other fields retain the same value as
in the type II deformed conifold solution \cite{ks}, modulo a trivial
rescaling of the R-R fields. 
We can now easily prove that all these assumptions are
compatible with the equations of motion. Consider the dilaton
and tachyon equations of motion in Einstein frame,
\begin{eqnarray}
\Box \phi &\sim& {1\over 2}e^{\phi}f_+(T)|F_{(3)}|^2-e^{-\phi}|H_{(3)}|^2+{T^2\over 2}e^{\phi/2}
\nonumber\\
\Box T &\sim& {1\over 2}e^{\phi}f^{\prime}_+(T)|F_{(3)}|^2+f^{\prime}_+(T)|F_{(5)}|^2+Te^{\phi/2}
\label{eq}
\end{eqnarray}
For $T\rightarrow 0$ the last term in the dilaton equation is negligible.
Since $f_+(0)=1$, the equation reduces to the type IIB dilaton
equation after a rescaling of the R-R fields. The
dilaton may remain constant. After neglecting subleading terms, there is 
a constant term
$e^{\phi}f_+(T)|F_{(3)}|^2\sim 1/(g_s^2M)$
in the right hand side of the equation of T. For the solution (\ref{IR}),
$\Box T\sim {1\over \tau^2}\partial_\tau (\tau^2\partial_\tau T)$.
The tachyon equation is then satisfied for $T\sim \tau^2$. 
We can similarly check that Einstein equations are still satisfied.

Alternatively, we can use the five-dimensional effective theory
(\ref{pot}). For $T=0$, a simple rescaling $(N,M,F)\rightarrow
\sqrt{2}(N,M,F)$ maps the type 0 potential to the type IIB one.
The change of variables $r\rightarrow \tau$ is useful.
It is then possible to systematically expand the potential and the
solution in power of $\tau^2$.

We expect that the gauge theory
coupling constant  blows up in the far IR. We can qualitatively identify
the behavior of the coupling constant by looking at the world-volume of a
D5$^+$ probe wrapped on the 2-cycle. The effective coupling constant is
\begin{equation} 
{1\over g(\tau)^2}=e^{\phi}k(T)\sqrt{g_{(2)}}\sqrt{1+|B|^2_{(2)}}
\label{coupl}
\end{equation}
where $k(T)$ testifies the coupling of the worldvolume theory 
to the tachyon field \cite{kt,garousi} and the metric and B-field 
are evaluated on the 2-cycle. Exactly as in \cite{ks},
the effective coupling constant (\ref{coupl}) will diverge due to the
metric and B-field on the 2-cycle, $1/g^2\sim \tau^2$
\footnote{ 
Notice that in the D3-branes type 0 solutions \cite{kt}
the dilaton necessarily runs. Here, we are saved by the 2-cycle.
Notice also some analogies.
An interesting solution in \cite{kt} flows to an IR fixed point 
at strong coupling. 
Such a particular flow exists also for the more general theories
$SU(Q)\times SU(N)$ of $N$ electric and $Q$ magnetic D3-branes \cite{big}.
In this (fine-tuned) solution, 
the tachyon goes to zero while the dilaton blows up
in a very precise combination,
with most of the fields retaining
the same value as in type II. In the conifold case, 
since the dilaton affects differently
the terms $F_{(5)}^2$ and  $F_{(3)}^2$, we have no
solution like that.} .

We can systematically expand the solution near the IR 
in power series of $\tau^2$.
We expect that, being regular, such solution describes a consistent
RG flow. We may conjecture 
that the previous asymptotic solution correctly  describes the IR
behavior of the pure $SU(M)$ theory.  However, 
the full solution describing the theory at 
all scales is certainly difficult to find.  
Since the solution involves a fine-tuning $T\rightarrow 0$, it is
not clear if it can really describe the pure $SU(M)$ theory decoupled
from its UV completion. We could speculate that $T=0$ is an IR attractor
for all solutions.
We envisage different possibilities for the UV theory. 
The IR solution can be connected via a cascade
mechanism to the  $SU(N+M)\times SU(N)$ theory with bi-fundamental
scalar fields, obtained with $N$ D3$^+$ and $M$ D5$^+$ branes. 
Alternatively, we could find a pure $SU(M)$ theory exhibiting 
asymptotic freedom in the UV,
as in \cite{kt}. 
A third option is a UV completion along the lines of \cite{mn}.
In all cases we face the problem of extrapolating our 
Lagrangian (\ref{lag}) to a region where $T$ is not small. The exact
form of $f_+(T)$ and of the potential $V(T)$ for T could become crucial 
for correctly capturing the UV behavior. 
Moreover the dilaton will start to run.
It would be quite interesting to see if the class of exact solutions
(\ref{black}) has a type 0 extension to the case of non-constant dilaton
and arbitrary $f_+(T),V(T)$.

\section{Conclusions}

We have seen that the deformed conifold solution can be easily adapted
to describe non-supersymmetric solutions. The use of a Calabi-Yau
manifold may suggest supersymmetry, but
the mechanism in \cite{ks} seems more general. The necessary
ingredients are RR fluxes supported by finite-size cycles. 
From eq. (\ref{eqZ}), we see that, in the absence of D3 sources
and with mild assumption on K, the warp factor Z is regular. 

These ingredients may define a larger class of solutions.
 We may expect to study interesting non-supersymmetric
theories by using generic manifolds.
Type 0 is extremely useful for constructing non-supersymmetric
gauge theories with branes.
However, in type 0, we are more restrained by the
absence of a sensible Lagrangian for generic T. It would be
interesting to extend the class of solutions (\ref{black}) to 
type 0. We noticed that, under the assumption that
T relaxes to zero, the solutions (\ref{black}) still provide
an IR asymptotic description of the physics. 

Moreover, in type 0 solutions, the stabilization of the tachyon
requires small t'Hooft coupling. Ultimately, this could be
not so dangerous, at least for our most ambitious goals.
The discovery of completely regular supergravity theories
can be useful for studying the dynamics of strongly coupled 
gauge theories.
We can typically describe with supergravity only gauge theories
with finite cut-off and coupled to extra-modes. 
Eventually, we will be interested in duals of weakly coupled theories.
As usual, we were prepared to the unpleasant situation that,
in the seek of a description of strong dynamics for 
weakly coupled gauge theories, we need to cross a region where
the effective t'Hooft parameter is small. This invalidates the
supergravity approximation and requires a string sigma-model
analysis.

\vskip .2in \noindent \textbf{Acknowledgments}\vskip .1in \noindent  
We would like to thank A. Armoni, S. Ferrara, A. Pasquinucci and A.
Tseytlin for 
useful discussions and suggestions.
L.G., F. B. and A. Z. are partially supported by INFN and MURST, and
by the European Commission TMR program HPRN-CT-2000-00131, 
wherein they are associated to the University of Padova.

\end{document}